\documentclass[twocolumn,pra,aps]{revtex4}
\usepackage{graphicx}

\def\duzomniejsze{<\kern-.7mm<}
\def\duzowieksze{>\kern-.7mm>}

\def\textbf#1{{\bf #1}}
\def\beq{\begin{equation}}
\def\eeq{\end{equation}}
\def\be{\begin{equation}}
\def\ee{\end{equation}}
\def\ben{\begin{eqnarray}}
\def\een{\end{eqnarray}}
\def\beqa{\begin{eqnarray}}
\def\eeqa{\end{eqnarray}}
\def\eea{\end{array}}
\def\bea{\begin{array}}
\newcommand{\bei}{\begin{itemize}}
\newcommand{\eei}{\end{itemize}}
\newcommand{\bee}{\begin{enumerate}}
\newcommand{\eee}{\end{enumerate}}

\def\>{\rangle}
\def\<{\langle}

\begin{document}

\title{Signature candidate of quantum chaos far from the semiclassical regime}

\begin{abstract}

We numerically investigated the entanglement product in the simplest
coupled kicked top model with the spin $j=1$. Different from the
dynamical pattern of entanglement in the semiclassical regime, two
similar initial states may have discordant entanglement
oscillations. A candidate of the quantum signature of this classical
chaotic system was proposed. The potential antimonotonic relation
between the rank correlation coefficient qualifying the concordant
of two entanglement evolutions and the stationary entanglement was
preliminarily revealed.

\end{abstract}
\author{Shang-Bin Li}, \author{Zhengyuan Xu}

\affiliation{School of Information Science and Technology, and
Optical Wireless Communication and Network Center, University of
Science and Technology of China, Hefei, Anhui 230027, P.R. China}
\maketitle

\textbf{The distinct dynamical behaviors of entanglement in the
simplest coupled kicked top model with the spin $j=1$ were found. It
was shown two similar initial states may have discordant
entanglement oscillations. Based on it, the rank correlation
coefficient qualifying the concordant of two entanglement evolutions
was proposed as the candidate of the quantum signature of this
classical chaotic system, which has the potential antimonotonic
relation with the stationary entanglement caused by a similar
dynamical process as quantum quench.}

Entanglement in the quantum chaotic systems has been studied by
several authors
\cite{zurek,h1,Ghose1,wang,sanders,stock,furuya,sarkar1,arul3,blprl4,tanaka,shepe1,shepe2,silves,shepely,angelo99,Li2006,abreu,trail,r,chung,kubotani,l,Y}.
For the system of weak coupled kicked tops, in the semiclassical
regime, it has been clarified that two initially separable
subsystems can get entangled in a nearly linear rate depending on
the intrinsic chaotic properties, and their entanglement eventually
reaches saturation \cite{sarkar1,arul3,blprl4,tanaka,r}. It has also
been elucidated that the increment of the nonlinear parameter of
weakly coupled kicked tops does not accelerate the entanglement
production in the strongly chaotic region \cite{tanaka}. For strong
coupled kicked tops in the semiclassical regime, it is observed the
greater the kick strength, the higher asymptotic value of
entanglement if the kick strength is not very small \cite{blprl4,r}.
In the deep quantum regime, based on the combined electronic and
nuclear spin of a single atom, the experimental realization of the
quantum kicked top with $j=3$ has been presented and good correspondence between the quantum dynamics and classical
phase space structures has been found \cite{Ghose1}. Clear differences in the
sensitivity to perturbation in chaotic versus regular regimes, and
experimental evidence for dynamical entanglement as a signature of
chaos have also been observed \cite{Ghose1}. In Ref.\cite{zurek02},
it has been shown that the discrepancy between quantum and classical
evolutions can be significantly decreased by even a weak loss of the
coherence. Motivated by previous works, we investigated the
entanglement quantified by log-negativity in coupled kicked tops
with the spin $j=1$ by using
$\rho=p|\psi_1\rangle\langle\psi_1|\otimes|\psi_2\rangle\langle\psi_2|+(1-p)|\psi_2\rangle\langle\psi_2|\otimes|\psi_1\rangle\langle\psi_1|$
as the initial state. The difference of the Von-Neumann entropy
between $\rho_{p=0}$ and $\rho_{p=0.5}$ is given by
$\Delta{S}=1-\frac{1}{\ln4}[(1-|\langle\psi_1|\psi_2\rangle|^2)\ln(1-|\langle\psi_1|\psi_2\rangle|^2)+(1+|\langle\psi_1|\psi_2\rangle|^2)\ln(1+|\langle\psi_1|\psi_2\rangle|^2)]$,
which depends on the overlap of $|\psi_1\rangle$ and
$|\psi_2\rangle$ and $\Delta{S}\in[0,1]$. The purpose of choosing
this kind of initial states is to keep the energy independent on the
initial mixedness for the following coupled symmetric kicked tops.
Moreover, this kind of initial states guarantees the
mixedness-independent dynamical trajectories of arbitrary symmetric
collective observables such as $\hat{J}^2_{z_1}+\hat{J}^2_{z_2}$ or
$\hat{J}_{z_1}\hat{J}_{z_2}$ in two coupled symmetric tops. It is
rationally conjectured that, in most cases, the characteristics of
the entanglement evolution $E_2(t)$ corresponding to the above
initial state with $p=0.5$ reflect closer details concerning its
classical counterpart than the entanglement evolution $E_1(t)$ with
initial $p=0$. Since in most of the processes of the quantum-classical transition, it comes with the loss or gain of the information \cite{zurek}. Good quantum-classical corresponding reflects
the robustness of the system to the change of the information entropy \cite{zurek}. Using the above specific initial states to elucidate the sensitivity of the dynamics to the initial change of no more than 1-qubit information in this simplest coupled kicked tops can help us to understand the quantum chaos in deep quantum regime. This is our main motivation. In order to investigate how the initially lost information
affects the entanglement dynamical behaviors of two tops with small
spin, it is proposed to apply the scaled rank correlation
coefficient \cite{kendallb} \be
\eta_d\equiv\lim_{N\rightarrow\infty}\frac{|\Delta{S}|}{N}\sum^{N}_{t=1}[\Theta({E}_1(t)-E_1(t-1))\Theta({E}_2(t)-E_2(t-1))]
\ee to quantify the concordant or discordant between two
entanglement evolution, where $\Theta(x>0)=1$, $\Theta(x=0)=0$ and
$\Theta(x<0)=-1$. The $\eta_d$ closer to 1 indicates two
entanglement evolutions are more concordant. For the weak coupled
kicked tops with $j>50$, previous studies have shown the
entanglement between two tops does not exhibit large amplitude
oscillation but monotonic ascent to saturation
\cite{sarkar1,arul3,blprl4,tanaka,r}. In these cases, $\eta_d$ is
always near one which can not provide the explicit quantum signature
of its classical chaotic counterpart. A cross correlation
coefficient $\eta_g$ is also defined for comparing with $\eta_d$ as
follows: \be
\eta_g\equiv\lim_{N\rightarrow\infty}\frac{1}{N}\sum^{N}_{t=1}[({E}_1(t)-E_1(t-1))({E}_2(t)-E_2(t-1))],
\ee which may also characterize the synchronization of the time
variations of $E_1(t)$ and $E_2(t)$.
\begin{figure}
\centerline{\includegraphics[width=3.5in]{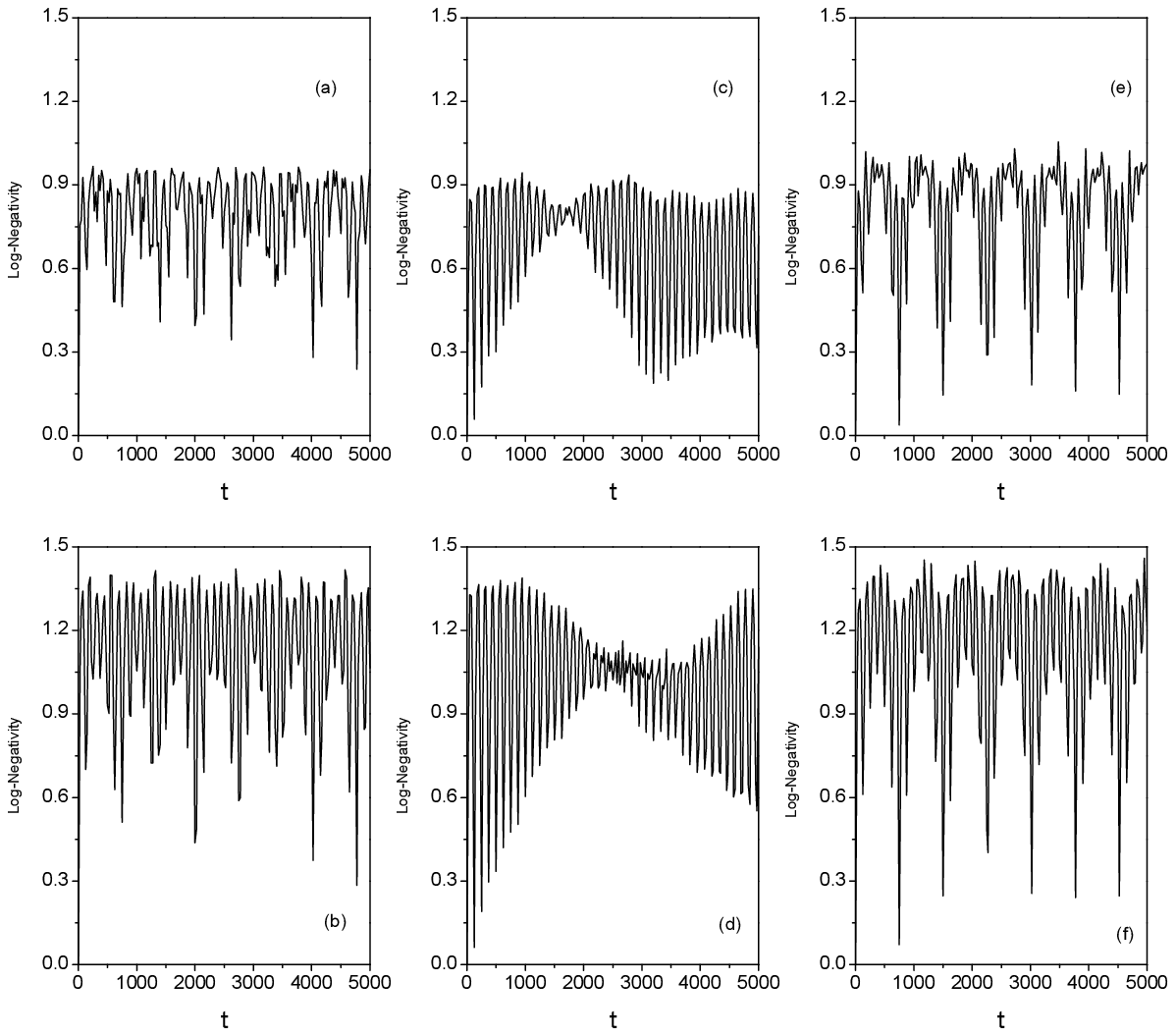}} \caption{The
log-negativity of two tops is plotted as the function of $t$ for
three different nonlinear coefficients and for pure or mixed spin
coherent state. $j=1$; $\varepsilon=0.05$; $\gamma_1=-3$;
$\gamma_2=3$; (a) $k=0.25$, $p=0.5$; (b) $k=0.25$, $p=0.0$; (c)
$k=3$, $p=0.5$; (d) $k=3$, $p=0.0$; (e) $k=6$, $p=0.5$; (f) $k=6$,
$p=0.0$.}
\end{figure}
\begin{figure}
\centerline{\includegraphics[width=3.5in]{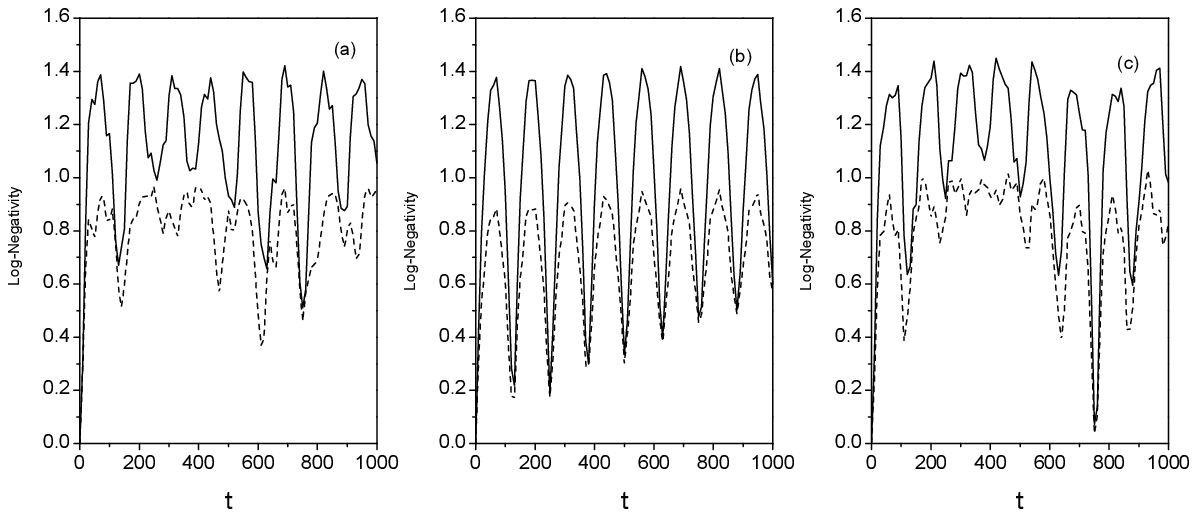}} \caption{The
same parameters as Fig.1; (a) $k=0.25$; (b) $k=3$; (c) $k=6$; (Solid
line) $p=0$; (Dash line) $p=0.5$. The scaled rank correlation
coefficients $\eta_d$ corresponding to $k$ have the order
relation $\eta_d(k=0.25)<\eta_d(k=6)<\eta_d(k=3)$. For other initial state with $\gamma_1=-\gamma_2=1$ (a fixed point in
phase space of the single kicked top with $k=3$ in
Ref.\cite{sarkar1}), we have also calculated the parameter $\eta_d$ and obtained $\eta_d(k=0.25)=0.314<\eta_d(k=6)=0.410<\eta_d(k=3)=0.448$.}
\end{figure}
\begin{figure}
\centerline{\includegraphics[width=2.5in]{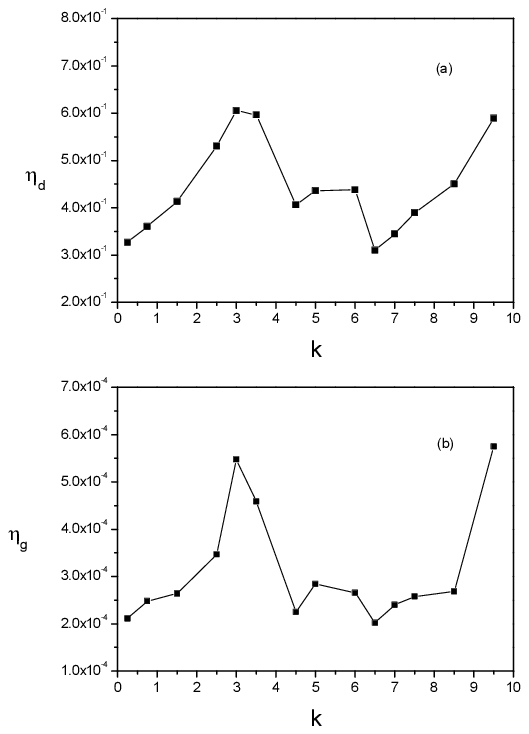}} \caption{The
correlation coefficients (a) $\eta_d$ and (b) $\eta_g$ are plotted as the function of $k$. $j=1$; $\varepsilon=0.05$; $\gamma_1=-3$;
$\gamma_2=3$.}
\end{figure}

The Hamiltonian of the coupled kicked tops can be written as
\cite{sarkar1}:
\begin{eqnarray}
{\cal \hat{H}}(t) &=& \hat{H}_1(t) + \hat{H}_2(t) + \hat{H}_{12}(t)\\
\mbox{with}~~\hat{H}_i(t) &\equiv& \frac{\pi}{2} \hat{J}_{y_i} +
\frac{k_i}{2j} \hat{J}_{z_i}^{2}
\sum_{n} \delta (t-n),\\
\hat{H}_{12}(t) &\equiv& \frac{\varepsilon}{j} \hat{J}_{z_1}
\hat{J}_{z_2} \sum_{n} \delta (t-n), \label{couple}
\end{eqnarray}
\noindent where $i = 1, 2$, and $\hat{H}_i(t)$'s are the
Hamiltonians of the individual tops. $\hat{H}_{12}(t)$ represents
the coupling between two tops with a coupling strength of
$\varepsilon/j$. All these spin operators obey standard commutation
relations. Choosing the simultaneous eigenstates $|j_1,
m_1\rangle_1\otimes|j_2, m_2 \rangle_2$ of the four mutually
commuting operators $\hat{J}_{i}^{2}$ and $\hat{J}_{z_i}$ ($i=1,2$)
as the basis, where $\hat{J}_{i}^{2} |j_i, m_i\rangle_i = j_i(j_i+1)
|j_i, m_i\rangle_i$ and $\hat{J}_{z_i} |j_i, m_i \rangle_i = m_i
|j_i, m_i\rangle_i$. Thereafter, it is assumed $j_1=j_2=j$ and
$k_1=k_2=k$ for the coupled kicked symmetrical tops.

In previous studies, usually the pure direct product state of two
spin coherent states \cite{sarkar1,arul3,blprl4,tanaka,r} is chosen
as the initial state of the total system. Here, we use a mixed state
of direct product of two spin coherent states as our initial state
for two tops. The spin coherent state is given in $|j, m_i
\rangle_i$ basis as : \be |\gamma\rangle_i = ( 1 + |\gamma|^2 )^{-j}
\sum^{j}_{m_i=-j}\gamma^{j - m_i} \sqrt{\left(\begin{array}{c} 2 j\\
j + m_i
\end{array} \right)}|j,m_i\rangle_i,
\ee where $\gamma \equiv \exp(i \phi) \tan (\theta/2)$. The initial
normalized mixed state can be written as \be
\rho(0)=p|\psi^{+}(0)\rangle\langle\psi^{+}(0)|+(1-p)|\psi^{-}(0)\rangle\langle\psi^{-}(0)|,
\ee where $|\psi^{+}(0)\rangle$ and $|\psi^{-}(0)\rangle$ are given
by: \beqa |\psi^{+}(0) \rangle &=&
|\gamma_1\rangle_1|\gamma_2\rangle_2 \nonumber\\
|\psi^{-}(0) \rangle &=&|\gamma_2\rangle_1|\gamma_1\rangle_2, \eeqa
and $p\in[0,1]$ is the real number.

The log-negativity of a bipartite density matrix is defined by
\cite{vidalwerner} \be {\mathcal{N}}(\rho)=\log_2\|\rho^{\Gamma}\|,
\ee where $\rho^{\Gamma}$ is the partial transpose of $\rho$ and
$\|\rho^{\Gamma}\|$ denotes the trace norm of $\rho^{\Gamma}$, which
is the sum of the singular values of $\rho^{\Gamma}$. The coupled
kicked tops with $j=1$ is far from the semiclassical regime. Due to
the exchange symmetry of ${\cal \hat{H}}(t)$ and its Floquet
operator $\hat{U}$, the Floquet eigenstates have well-defined
exchange parity: $j_e=(j+1)(2j+1)$ Floquet eigenstates $|e_i\rangle$
with even exchange parity and $j_o=j(2j+1)$ Floquet eigenstates
$|o_i\rangle$ with odd exchange parity. The Floquet operator
$\hat{U}$ can be expressed as \be
\hat{U}=\sum^{j_e}_{i=1}e^{\imath\phi_i}|e_i\rangle\langle{e}_i|+\sum^{j_e+j_o}_{i=j_e+1}e^{\imath\phi_i}|o_i\rangle\langle{o}_i|.\ee
The initial state can be expanded by the Floquet states as \be
|\gamma_2\rangle_1|\gamma_1\rangle_2=\sum^{j_e}_{i=1}a_i|e_i\rangle+\sum^{j_e+j_o}_{i=j_e+1}b_i|o_i\rangle,\ee
and \be
|\gamma_1\rangle_1|\gamma_2\rangle_2=\sum^{j_e}_{i=1}a_i|e_i\rangle-\sum^{j_e+j_o}_{i=j_e+1}b_i|o_i\rangle,\ee
where $a_i$ and $b_i$ are the complex coefficients. The evolving
state corresponding to the initial states $\rho_{p=0}(0)$ and
$\rho_{p=0.5}(0)$ can be obtained as \be
\hat{U}^n|\gamma_2\rangle_1|\gamma_1\rangle_2=\sum^{j_e}_{i=1}a_ie^{\imath{n}\phi_i}|e_i\rangle+\sum^{j_e+j_o}_{i=j_e+1}b_ie^{\imath{n}\phi_i}|o_i\rangle,\ee
and \beqa
\hat{U}^n\rho_{p=0.5}(0)\hat{U}^{\dagger{n}}&&=\sum^{j_e}_{i,l=1}a_ia^{\ast}_le^{\imath{n}(\phi_i-\phi_l)}|e_i\rangle\langle{e}_l|\nonumber\\
&&+\sum^{j_e+j_o}_{i,l=j_e+1}b_ib^{\ast}_le^{\imath{n}(\phi_i-\phi_l)}|o_i\rangle\langle{o}_l|,\eeqa
respectively. Comparing the above two evolving states, it is
explicit that the scaled rank correlation coefficient $\eta_d$
reflects the influence of coherent superposition between two Hilbert
subspaces with different exchange parity on the dynamical behaviors
of entanglement. In Figs.(1-2), we can see the log-negativity of the
time evolution density operator for two different mixedness and for
$k=6$, $k=3$ and $k=0.25$, which are strong chaotic, weak chaotic
and regular cases of the single kicked top in previous literatures
\cite{dariano}, respectively. For all of the cases, we pick an
initial state in Eqs.(7-8) with $\gamma_1=-3$ and $\gamma_2=3$.
Slightly different from the results in Ref.\cite{Ghose1}, the
log-negativity shows quasi-periodic behavior and collapses and
revivals of the oscillation for $k=3$. While for the case of $k=6$,
the global dynamical pattern of the log-negativity is quasi-periodic
but local dynamical pattern is chaotic. For small value of $k=0.25$,
i.e. the regular case, the log-negativity only exhibits a chaotic
oscillation. For the mixed initial state in Eqs.(7-8), the evolving
log-negativity with $p=0.5$ is the lower bound of the one with
$p=0$. The scaled rank correlation coefficient $\eta_d$
quantifying the concordant or discordant of the entanglement
evolutions corresponding to initial states with $p=0$ and $p=0.5$
depends on $k$ even though the corresponding evolving states contain
the same dynamics for arbitrary symmetric collective operators such
as $\hat{J}^2_{z_1}+\hat{J}^2_{z_2}$ or the coupling term
$\hat{J}_{z_1}\hat{J}_{z_2}$. From Fig.2(b), it can be found that
the corresponding entanglement dynamical behaviors of two cases with
$p=0$ and $p=0.5$ are synchronous and ordered when $k=3$, not like
the disorder and asychronous correspondence in Fig.2(a) and
Fig.2(c). In Fig.3, we plot $\eta_d$ and $\eta_g$ as the functions of $k$. It is shown both $\eta_d$ and $\eta_g$ firstly
increase from very small value of $k$ and achieve their local maximal values near $k=3$, then non-monotonically
decrease to their local minimal values near $k=6.6$, and subsequently revival. For relating the scaled rank correlation coefficient $\eta_d$ in coupled kicked tops with $j=1$, $k=3$ and $\varepsilon=0.05$
with the previous analysis about its chaotic behaviors of classical counterpart, it is desirable to discuss the initial states dependence of $\eta_d$. For example, choosing four cases with
$\gamma_1=-\gamma_2=1,\tan(\frac{2.25}{2})e^{0.63i},3,\tan(\frac{0.89}{2})e^{0.63i}$ (fixed, fixed, chaotic, chaotic points in
phase space of the single kicked top with $k=3$ in
Ref.\cite{sarkar1}, respectively), the corresponding scaled rank correlation coefficients $\eta_d$ could be calculated and the results are $\eta_d=0.448, 0.468, 0.605, 0.504$, respectively, which implies the more classical chaotic initial state of individual top will induce the more synchronous dynamical behaviors of $E_1(t)$ and $E_2(t)$.

In Ref.\cite{Ghose1}, it has been experimentally demonstrated that
less average entanglement generation for initial states localized in
regular regions compared to those in the chaotic sea. Similarly,
taking into account of the external pulse monitoring the collective
observable $\hat{J}^2_{z_1}+\hat{J}^2_{z_2}$ of two spins which
makes the states of two spins collapse toward the subspace with
${}_i\langle\hat{J}^2_{z_i}\rangle_i=j^2$ ($i=1,2$, and
${}_i\langle\hat{Y}\rangle_i$ denotes the expectation value of the
operator $\hat{Y}$), we investigated the entanglement production in
the coupled nonhermitian kicked tops by adding a positive imaginary
part of $k$ in the Hamiltonian in Eq.(4). The results are shown in
Figs.(4-6). Both the log-negativity of the tops with $k=0.25$ and
$k=6$ exhibit the damped chaotic oscillation and converge to be
stationary. The log-negativity at $t=1000$ firstly decreases with
the increase of ${\mathrm{Re}}(k)$ from $0$ to $3$, and reaches a
flat bottom, then increases again and achieves a local maximal value
near $6.4$. For further increase of ${\mathrm{Re}}(k)$, it was
numerically verified similar cycles repeat, and the stationary
log-negativity is about $0.14$ at $k=9$, where the stationary
log-negativity is defined as ${\mathcal{N}}(t\rightarrow\infty)$. It
has been verified the stationary log-negativity is independent of
the initial pure or mixed states defined by Eqs.(7-8). From
Fig.6(a), in the case with ${\mathrm{Re}}(k)=3$, it can be seen that
the stationary log-negativity approximately linearly increases with
the coupling strength $\varepsilon$ and the influence of
${\mathrm{Im}}(k)$ is negligible when $\varepsilon\leq0.1$. Near
$\varepsilon=1$, the stationary log-negativity achieves its local
maximal values. Surprisingly, the stationary log-negativity could
approach to 1.55, which implies, independent of the initial pure or
mixed states, the stationary state is very close to the maximally
entangled qutrits in the simplest non-Hermitian coupled kicked tops.
Further increase of the coupling strength will make two coupled
kicked tops very difficultly evolve into the stationary state even
with a very large imaginary part of $k$. Their entanglement becomes
stationary after at least $10^5$ kicking steps. While for the case
with ${\mathrm{Re}}(k)=0.25$, the imaginary part of $k$ heavily
affects the stationary entanglement, and the results were shown in
Fig.6(b). Similarly, the stationary log-negativity has its local
maximum at certain value of $\varepsilon$ between $0.7$ and $0.8$,
and declines to 1.

Similar to
the previous results that chaotic (regular) semiclassical dynamical behaviors have connections with large (small)
entanglement production in their corresponding quantum systems in Ref.\cite{Ghose1,wang,sanders,stock,r},
from the Fig.2 and Fig.5, it can be found the
larger (smaller) stationary log-negativity, the more chaotic
(regular) dynamical behaviors of the log-negativity in the present quantum system with only total 9-dimensional Hilbert space. Moreover, comparing Fig.3(a) and Fig.5, the antimonotonic relation
between the rank correlation coefficient and the stationary entanglement can be found, in which the mismatch over a short range between $k=4.5$ and $k=6$
could be conjecturally removed via taking average over a large number of initial states.
Thus it is proposed to use this stationary log-negativity or $\eta_d$ defined
in Eq.(1) as a candidate of the signature of quantum chaos in deep
quantum regime. Obviously, the strength of the imaginary part
${\mathrm{Im}}(k)$ will affect the dynamical behaviors of the
log-negativity. If the stationary log-negativity was adopted as the
signature to investigate the border between the regular and chaotic
regions in the stroboscopic quantum phase space or parameter space,
it is suggested to choose a small enough imaginary part of $k$, i.e.
${\mathrm{Im}}(k)\ll{\mathrm{Re}}(k)$ to avoid significant
disturbance to the system of interest.
\begin{figure}
\centerline{\includegraphics[width=2.0in]{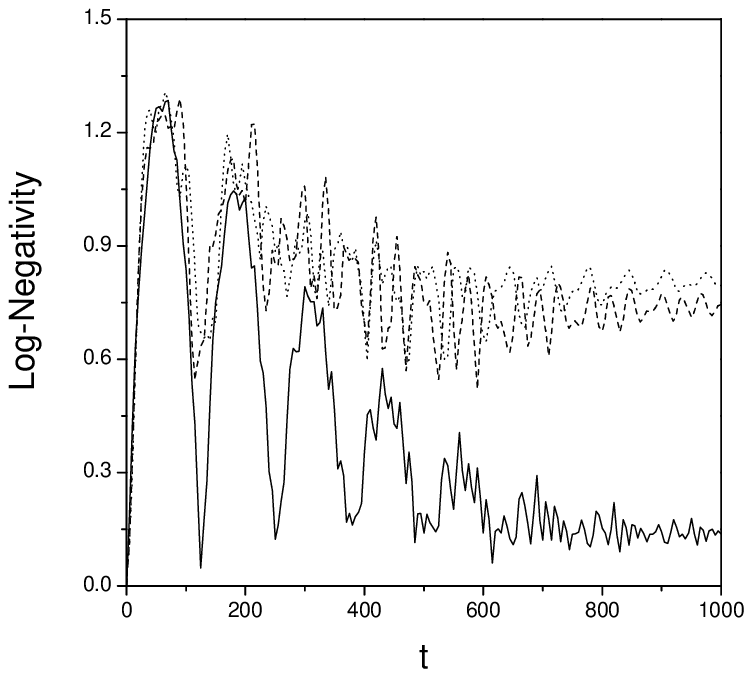}} \caption{The
log-negativity of two tops is plotted as the function of $t$ for
different nonlinear coefficients and pure initial spin coherent
state. $j=1$; $\varepsilon=0.05$; $\gamma_1=-3$; $\gamma_2=3$;
$p=0.0$; ${\mathrm{Im}}(k)=0.01$; (Solid line)
${\mathrm{Re}}(k)=3.0$; (Dash line) ${\mathrm{Re}}(k)=6.0$; (Dot
line) ${\mathrm{Re}}(k)=0.25$.}
\end{figure}
\begin{figure}
\centerline{\includegraphics[width=2.0in]{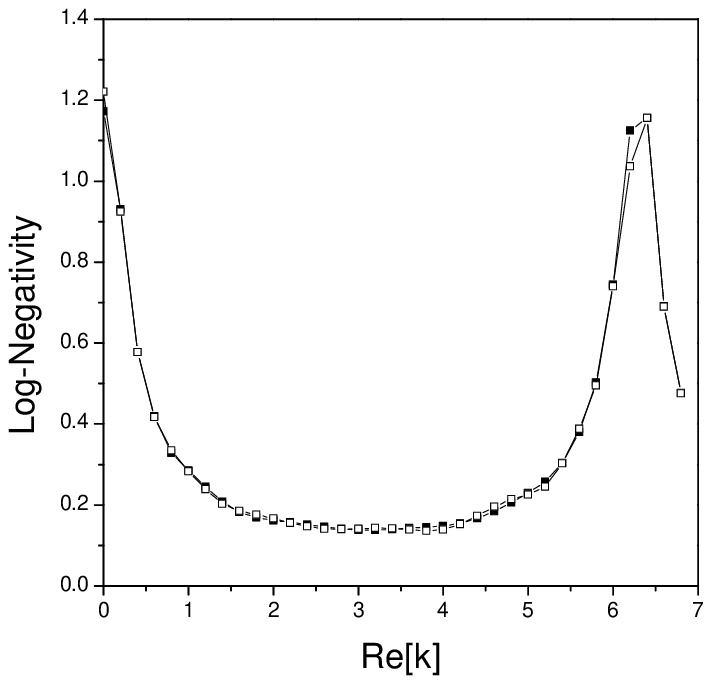}} \caption{The
log-negativity of two tops at the kick step $t=1000$ is plotted as
the function of the real part ${\mathrm{Re}}(k)$ of $k$ for two
different pure initial spin coherent state. $j=1$;
$\varepsilon=0.05$; $p=0$; ${\mathrm{Im}}(k)=0.01$; (hole square)
$\gamma_1=\gamma_2=\exp(0.63i)\tan(0.89/2)$, a chaotic point in
phase space of the single kicked top with $k=3$ in
Ref.\cite{sarkar1}; (Real square) $\gamma_1=-3$; $\gamma_2=3$.}
\end{figure}
\begin{figure}
\centerline{\includegraphics[width=2.0in]{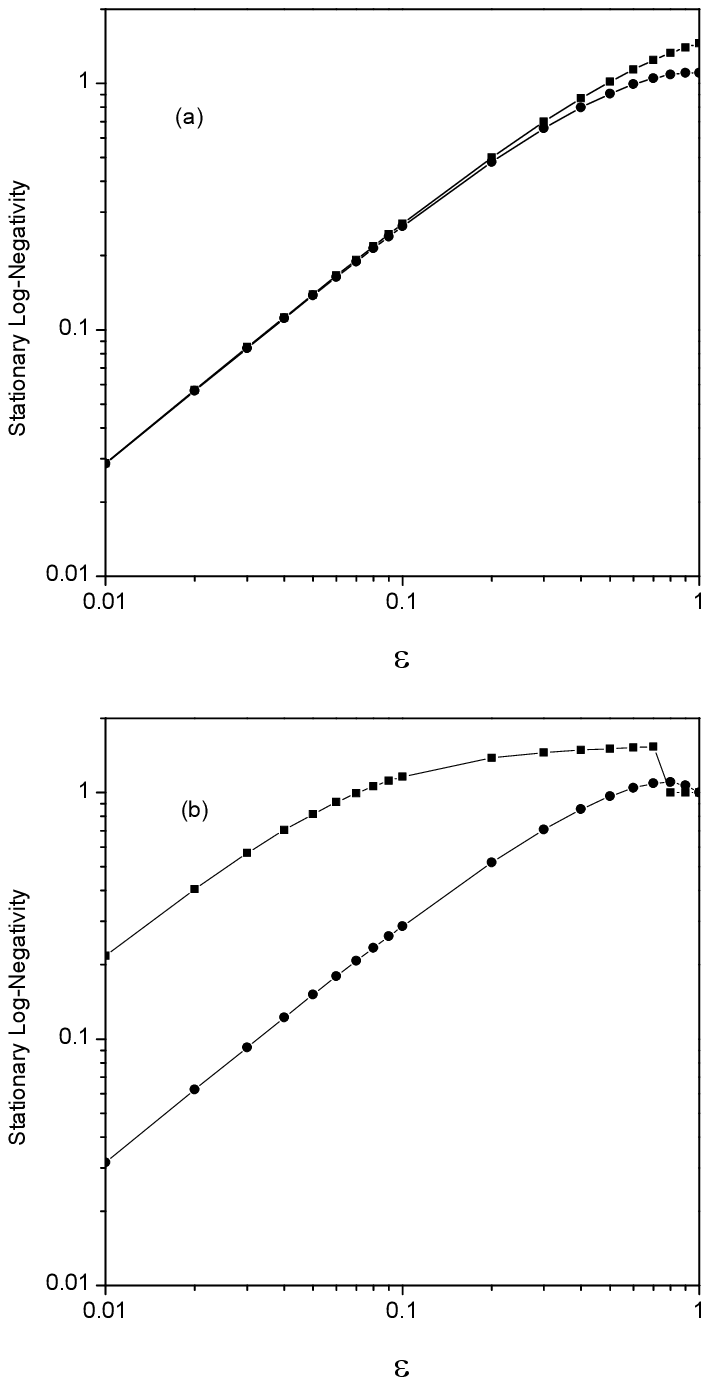}} \caption{The
stationary log-negativity of two tops is plotted as the function of
the coupling strength $\varepsilon$. $j=1$. (a)
${\mathrm{Re}}(k)=3$; (b) ${\mathrm{Re}}(k)=0.25$. (Square)
${\mathrm{Im}}(k)=0.01$; (Circle) ${\mathrm{Im}}(k)=3.0$. In the
case with $j=1$, it has been verified the stationary log-negativity
is initial-state independent for both pure or mixed initial states
in Eq.(7). In the case with ${\mathrm{Re}}(k)=3$,
${\mathcal{N}}(t\rightarrow\infty)\propto\varepsilon^{0.97}$ as
$\varepsilon\leq0.1$.}
\end{figure}

In Fig.2, it can be observed the log-negativity of two tops achieves
a local maximal value near $t=70$ for different values of the
parameters $k$ and $p$. We have investigated the dependence of the
local maximal log-negativity on the parameter $p$ for different
values of $k$ and a fixed value of $\varepsilon=0.05$. They
decline with $p$ at the slope of about 0.9 for the case with
$\gamma_1=-3$ and $\gamma_2=3$. The slopes only slightly depend on
the parameter $k$, but heavily correlate with the initial points
labeled by $\gamma_1$ and $\gamma_2$. In the cases of
$\gamma_1=-\gamma_2$, it is found the slopes decrease with the
increase of absolute value of $\gamma_1$ due to the mixedness of the
initial state decreases with $|\gamma_1|\geq1$ for $p\in(0,0.5]$.
For those kind of initial states in Eqs.(7-8), there may be quantum
correlation between two tops though they are separable
\cite{zurek06,vedral}. Their quantum discord is dependent of the
overlap $\langle\gamma_1|\gamma_2\rangle$ and $p$. The intermediate
value of the overlap and $p$ could make two tops achieve the larger
quantum discord. It is also interesting to investigate how the
quantum discord and classical correlation evolve in the kicked
coupled tops, which maybe embody good quantum-classical
correspondence \cite{zurekla,haroche,bli,diana}.
\begin{figure}
\centerline{\includegraphics[width=2.0in]{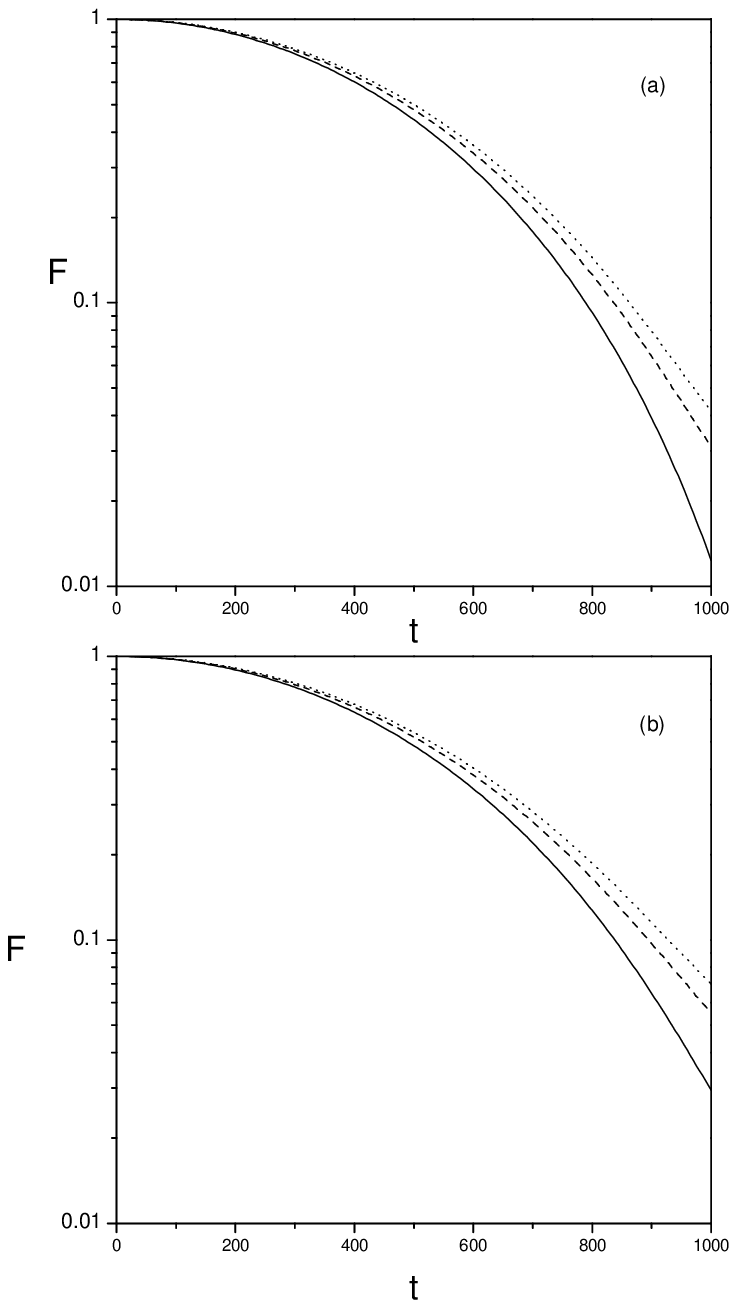}} \caption{The
fidelity of the system is plotted as the function of $t$ with $j=1$,
$k=0.25$, $k^{\prime}=0.26$, $\gamma_1=-3$, $\gamma_2=3$. (a)
$\varepsilon/j=0.02$, (Solid line) $p=0$; (Dash line) $p=0.2$; (Dot
line) $p=0.5$. (b) $\varepsilon/j=0.05$, (Solid line) $p=0$; (Dash
line) $p=0.2$; (Dot line) $p=0.5$. }
\end{figure}
\begin{figure}
\centerline{\includegraphics[width=2.0in]{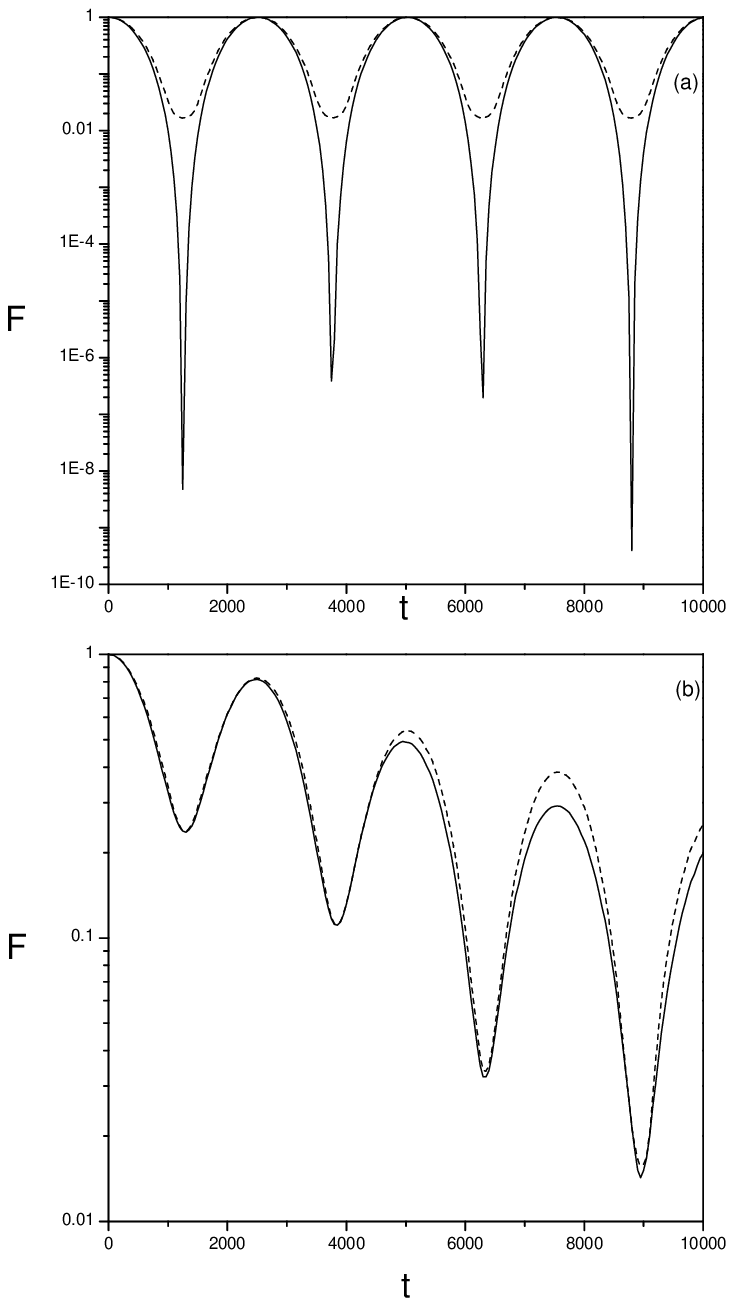}} \caption{The
fidelity of the system is plotted as the function of $t$ with $j=1$,
$k=0.25$, $k^{\prime}=0.26$, $\gamma_1=-3$, $\gamma_2=3$. (a)
$\varepsilon/j=0$, (Solid line) $p=0$; (Dash line) $p=0.5$. (b)
$\varepsilon/j=0.5$, (Solid line) $p=0$; (Dash line) $p=0.5$. }
\end{figure}

At last but not least, we should investigate the fidelity decay in the
coupled kicked tops. Usually, the fidelity for two mixed states
$\rho_1$ and $\rho_2$ is defined by \cite{joshi}\be
F(\rho_1,\rho_2)=[{\mathrm{Tr}}(\sqrt{\sqrt{\rho_1}\rho_2\sqrt{\rho_1}})]^2
\ee Here, we adopt this fidelity to study the sensitivity of
evolving state in the coupled kicked tops to the small perturbation,
i.e. $\rho_1(t)$ and $\rho_2(t)$ represent two evolving density
matrix of the kicked coupled tops with two slight different
nonlinear coefficients $k$ and $k^{\prime}=k+\delta{k}$
($\delta{k}\ll1$), respectively. Initially $\rho_1(0)=\rho_2(0)$
belong to those class of states in Eqs.(7-8). Generally, the
fidelity decay should be taken average of the overall initial states
in the phase space, and the average fidelity decay rate can be used to build
relations with the signatures of its classical chaotic counterpart
such as the mean Lyapunov exponent \cite{peres,fidelity,j,E}. The
average fidelity decay typically exhibits three consecutive stages,
i.e. the short-time parabolic decay, the followed intermediate-time
asymptotic decay, and long-time saturation \cite{fidelity,j,P}.
Nevertheless, it is still worthy to analyze the corresponding
fidelity decay for some specific evolving states with manifested
chaotic dynamical behaviors of entanglement. The key parameters in
Fig.(2a) are adopted to reveal how the fidelity decays in such a case
with chaotic dynamical behavior of bipartite entanglement. The
results are displayed in Figs.(7-8). In the first stage, the
fidelity more rapidly decays than exponential decay, and the decay
rate decreases with the increase of mixedness and the coupling
coefficient $\varepsilon$. Then, the fidelity revivals. When
$\varepsilon=0$, it can be observed perfect recurrence of the
fidelity, which implies the uncoupled kicked top with $k=0.25$ is
regular in the long-time scale. In the strong coupling case with
$\varepsilon=0.5$, initially the two trajectories of fidelity decay
seem coincident, but after some time, the divergence is obvious and
enlarged. In the kicked coupled spin-1, the coupling coefficient
$\varepsilon$ plays the more crucial role in the chaotic dynamical
behaviors of the entanglement than the nonlinear coefficient $k$, which may be the partial reason why
the classical chaotic initial states of the individual top induces the more synchronous dynamical behaviors of entanglement than the regular initial states.
Via replacing the operators of the spin $j=1$ by
$\hat{J}^{(i)}_z=\sum^{2}_{n=1}\sigma^{(i)}_{z_n}/2$ and
$\hat{J}^{(i)}_y=\sum^{2}_{n=1}\sigma^{(i)}_{y_n}/2$, where
$\sigma^{(i)}_{z_n}$ and $\sigma^{(i)}_{y_n}$ are the Pauli
operators of the qubit with label $\{i=1,2;n=1,2\}$, the kicked
coupled spin 1 is equivalent to a kicked four-qubit 2D Ising square.
Four-qubit 2D Ising square may be one of the basic cells for
constructing the universal quantum gate or measurement-based quantum
computation \cite{briegel}. Therefore, to fully investigate the
border between the regular and chaotic regions both in the quantum
phase space or parameter space is very important and desirable in
the future work.

In summary, this Letter deals with the dynamical pattern of
entanglement in the simplest kicked coupled symmetric tops with
$j=1$. The main result is to preliminarily reveal the potential
antimonotonic relation between the scaled rank correlation
coefficient $\eta_d$ defined in Eq.(1) and the stationary
entanglement induced by suitable nonhermitian term in the coupled
kicked tops. However, further calculations including large number of
initial states and key parameters in the system of the coupled
kicked tops are necessary to finally qualify these signature
candidates of quantum chaos in low dimensional Hilbert space.
Meanwhile, there is still lack of a reasonable physical
interpretation why the rank correlation coefficient has such an
antimonotonic relation with the stationary entanglement in the
system of amplified kicked coupled spin-1.

This work was supported in part by National 973 Program of China (Grant No. 2013CB329201).

\end{document}